\documentclass[prb,twocolumn,superscriptaddress,showpacs,groupedaddress,nofootinbib]{revtex4-1}

\usepackage{amsxtra}			% for boldsymbol
\usepackage{bbold}                  % for identity symbol
\usepackage[percent]{overpic} 	% for figure overlays

\usepackage[
pdfpagelabels, 
bookmarks = true,
plainpages=false,
bookmarksopen = true,
bookmarksnumbered = true,
breaklinks = true,
hypertexnames=true,
hidelinks=true,
linkcolor = black,
urlcolor  = red,
citecolor = black,
anchorcolor = black,
unicode,
hyperindex = true,
hyperfigures,
]{hyperref}

\newcommand{\eq}[1]{Eq.~\ref{#1}}
\newcommand{\fig}[1]{Fig.~\ref{#1}}
\newcommand{\refr}[1]{Ref.~\onlinecite{#1}}
\renewcommand{\exp}{\mbox{e}}
\newcommand{\rhoT}{\rho_{\mbox{\tiny T}}}
\newcommand{\rhoR}{\rho_{\mbox{\tiny R}}}
\newcommand{\rhoSpin}{\rho_{\mbox{\tiny S}}}

\newcommand{\ktot}{k^{\mbox{\tiny tot.}}}
\newcommand{\deltakx}{\boldsymbol \Delta_{1}}
\newcommand{\deltaky}{\boldsymbol \Delta_{2}}
\newcommand{\deltamb}{\Delta_{\mbox{\tiny MB}}}

\graphicspath{{figures}{./figures/}}

\newcommand{\placeFigMagnetisation}
{
\begin{figure}[t]
\begin{minipage}{\columnwidth}
\hspace{-2mm}
\begin{overpic}[width=0.5\textwidth]{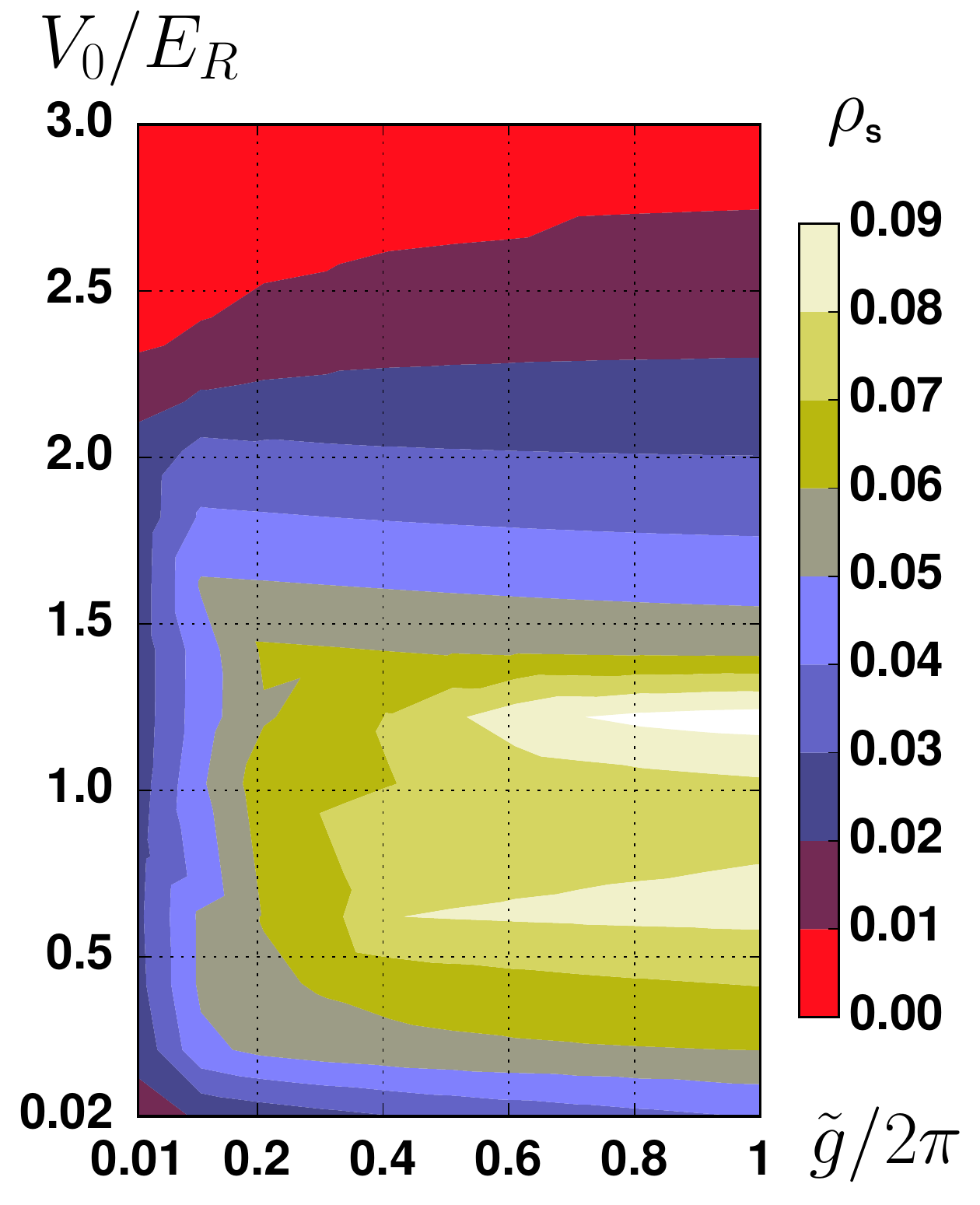}
\put(35,95){(a)}
\end{overpic}
\hspace{-2mm}
\begin{overpic}[width= 0.5\textwidth]{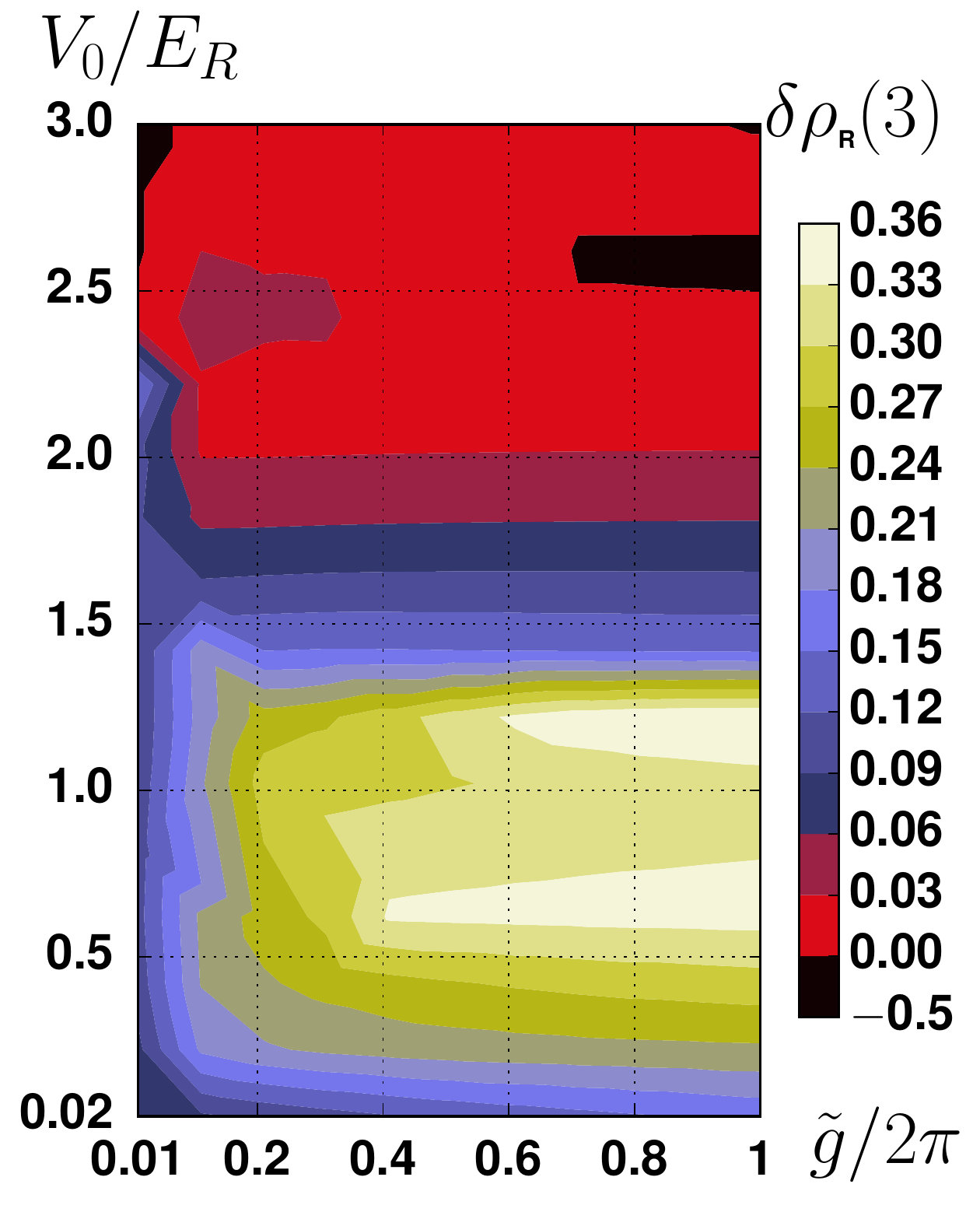}
\put(35,95){(b)}
\end{overpic}

\end{minipage}
\caption{
(color online). 
Map of the ground state character of $H_{\mbox{\tiny lb}}$ (\eq{eqInteractingHamiltonian}) as a function of lattice depth $V_0$ and interaction strength $\tilde g$ at $\nu=1/4$ [$N=9$ on a $6\times6$ simulation grid]:
(a) Spin-weighted density-density function $\rhoSpin$ [defined in the text], indicating a (partially) spin-polarized ground state where $1>\rhoSpin>0$. 
(b) Fluctuation $\delta \rhoR (3)$ of $\rhoR (3)$ from its mean value, as defined in \eq{eqFluctuationsR}, indicating breaking of 6-fold to 3-fold rotational symmetry where $\delta \rhoR (3) \sim 1$.}
\label{figMagneticPhaseDiagram}
\end{figure}
}

\newcommand{\placeFigCompressibility}
{
\begin{figure}[t]
\includegraphics[width=\columnwidth]{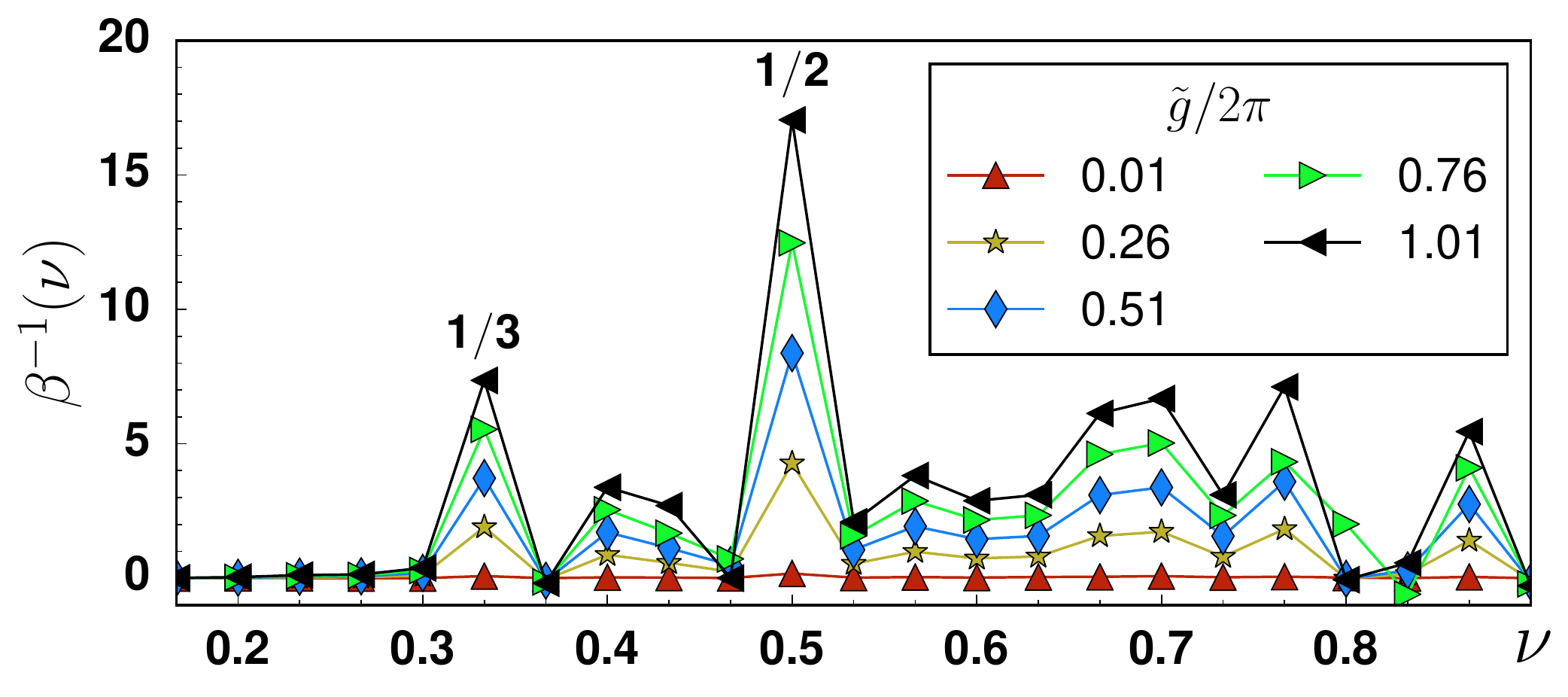}
\caption{
(color online). Inverse compressibility $\beta^{-1}(\nu)$ [defined in the text] for the ground state of $H_{\mbox{\tiny lb}}$ (\eq{eqInteractingHamiltonian}) as a function of filling factor $\nu$ of the lowest OFL band, for different values of interaction strength $\tilde g$ in the narrow band regime. Calculation for a $5\times6$ grid with $V_0/E_R =2$. Highly incompressible states occur at $\nu=1/3$ and $\nu=1/2$ once interactions are present.
}
\label{figIncompressibility}
\end{figure}
}

\newcommand{\placeFigLaughlinGroundState}
{
\begin{figure}[t]
\includegraphics[width=\columnwidth]{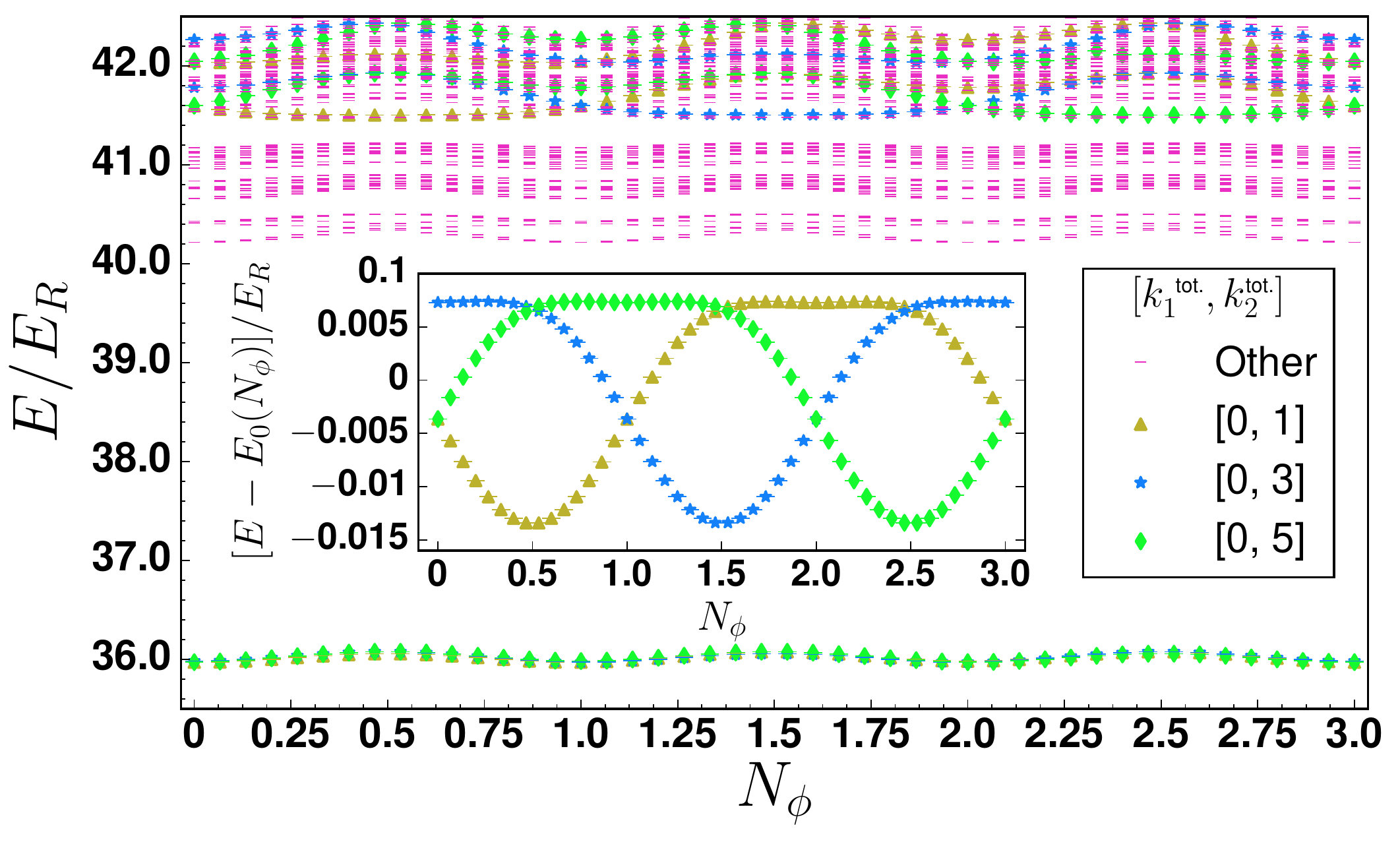}
\caption{
(color online). Energy spectrum of $H_{\mbox{\tiny lb}}$ (\eq{eqInteractingHamiltonian}) for filling factor $\nu=1/3$ of the lowest OFL band [$N=10$ on a $5\times6$ grid] with $V_0/E_R=2.0$ and $\tilde g/2\pi = 1$ as a function the number, $N_{\phi}$, of flux quanta inserted along the $\boldsymbol \kappa'_2$ direction ($N_{\phi}=1$ corresponds to $\deltaky = \kappa'_2/L_2$). Inset: Magnification of the ground state manifold over the same region with the mean ground state energy, $E_0(N_{\phi})$, subtracted off. The 3 ground states exhibit spectral flow whereby they return to their original ordering only after 3 flux quanta are inserted. 
}
\label{figLaughlinGroundState}
\end{figure}
}

\newcommand{\placeFigLaughlinCorrelations}
{
\begin{figure}[t]
\begin{minipage}{\columnwidth}
\hspace{-2mm}
\begin{overpic}[width= 0.5\textwidth]{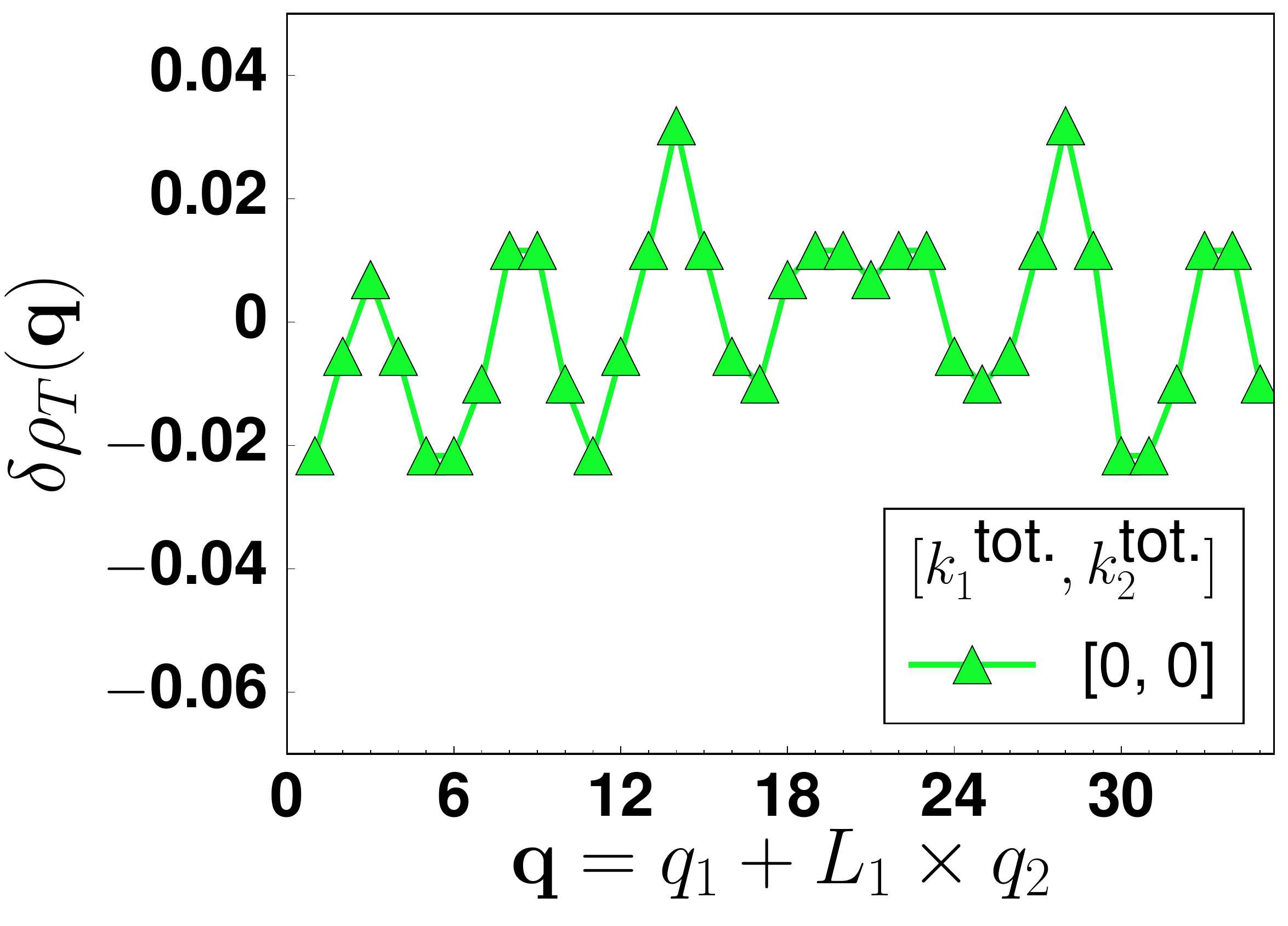}
\put(5,0){(a)}
\end{overpic}
\begin{overpic}[width= 0.5\textwidth]{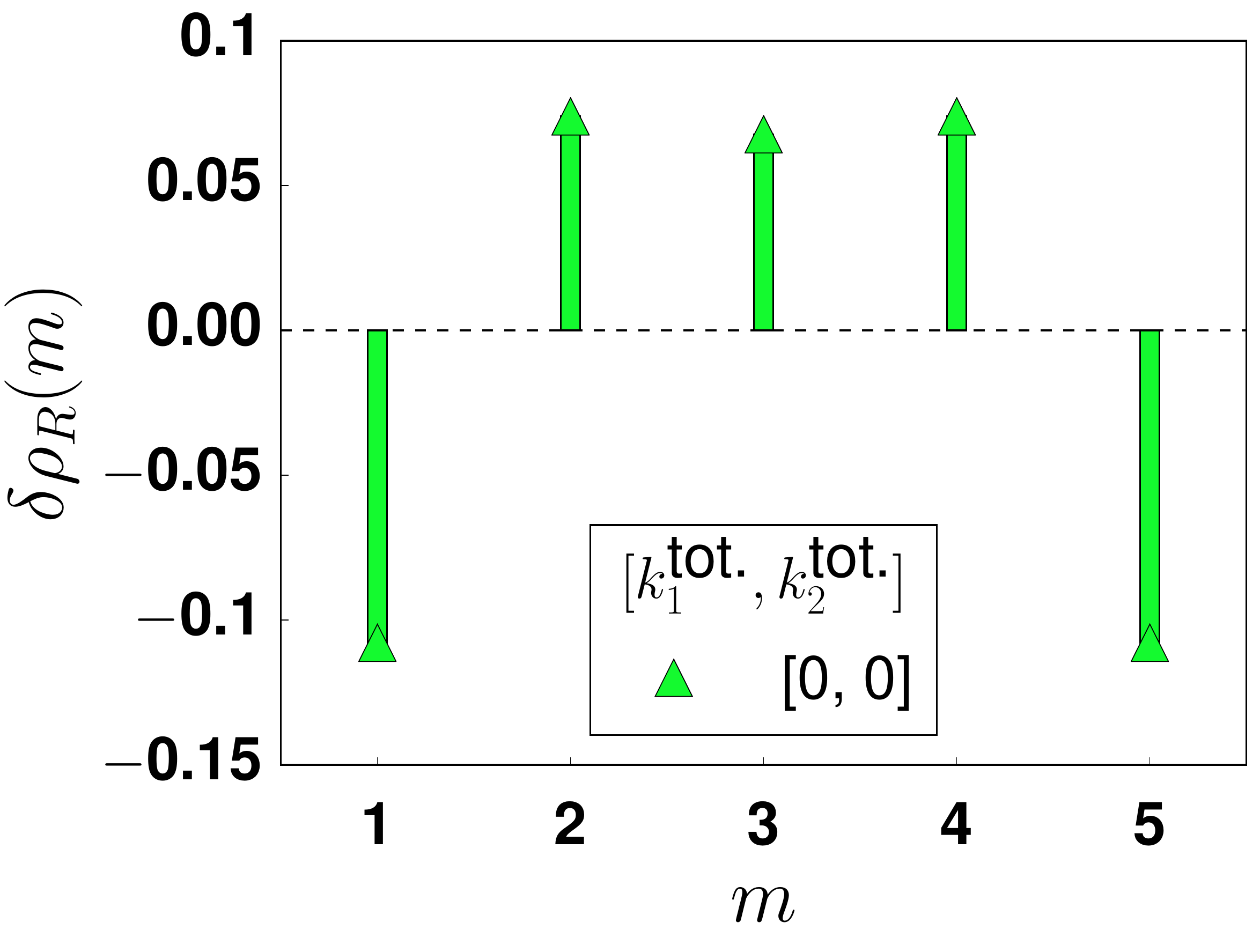}
\put(5,0){(b)}
\end{overpic}
\end{minipage}
\caption{
(color online). Ground state character of $H_{\mbox{\tiny lb}}$ (\eq{eqInteractingHamiltonian}) for filling factor $\nu=1/3$ of the lowest OFL band [$N=12$ on a $6\times6$ grid]: (a) Representative fluctuations $\delta\rhoT ({\mathbf q}) $ in the static structure factor $\rhoT ({\mathbf q}) $ [defined in the text]. $\left|\delta\rhoT ({\mathbf q}) \right| \sim 1/N $ here indicates translational invariance; (b) Representative fluctuations $\delta\rhoR (m) $ in the rotational correlation function $\rhoR (m)$ [defined in the text]. $\left|\delta\rhoR (m) \right| \sim 1/N $ here indicates 6-fold rotational invariance. 
}
\label{figLaughlinCorrelations}
\end{figure}
}

\newcommand{\placeFigParticipationRatio}
{
\begin{figure}[t]
\begin{minipage}{\columnwidth}
\begin{overpic}[width=0.9\columnwidth]{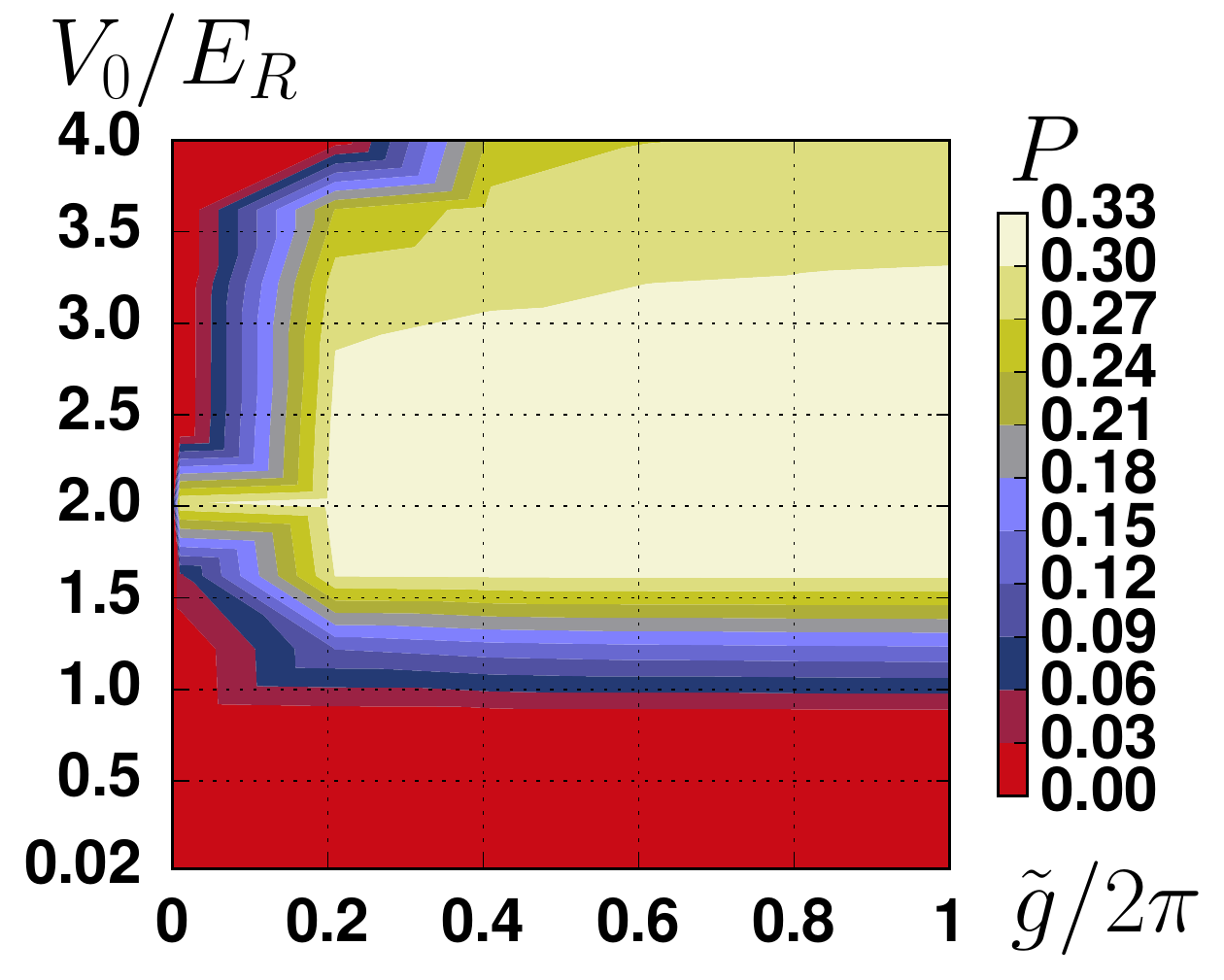}
\end{overpic}
\end{minipage}
\caption{
(color online). Fractional participation ratio $P$ (defined in the text), estimating the fraction of Fock states contributing to the ground state of $H_{\mbox{\tiny lb}}$ (\eq{eqInteractingHamiltonian}), for a range of lattice depth and interaction strength for $N=10$ on a $5\times6$ grid ($\nu=1/3$). 
}
\label{figParticipationRatio}
\end{figure}
}

\newcommand{\placeFigGapExtrapolation}
{
\begin{figure}[t]
\begin{minipage}{\columnwidth}
\hspace{-2mm}
\begin{overpic}[width=0.5\textwidth]{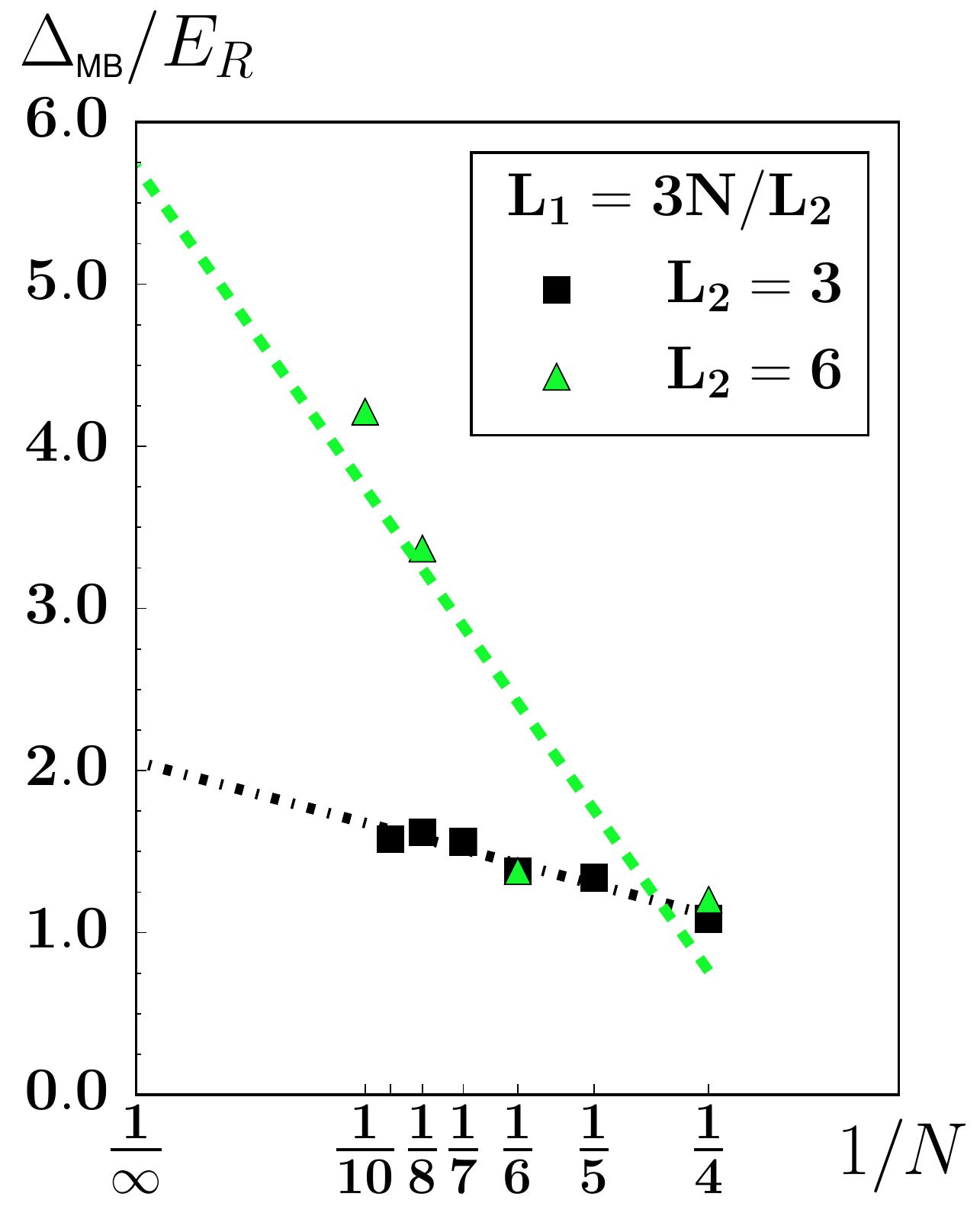}
\put(35,95){(a)}
\end{overpic}
\hspace{-2mm}
\begin{overpic}[width= 0.5\textwidth]{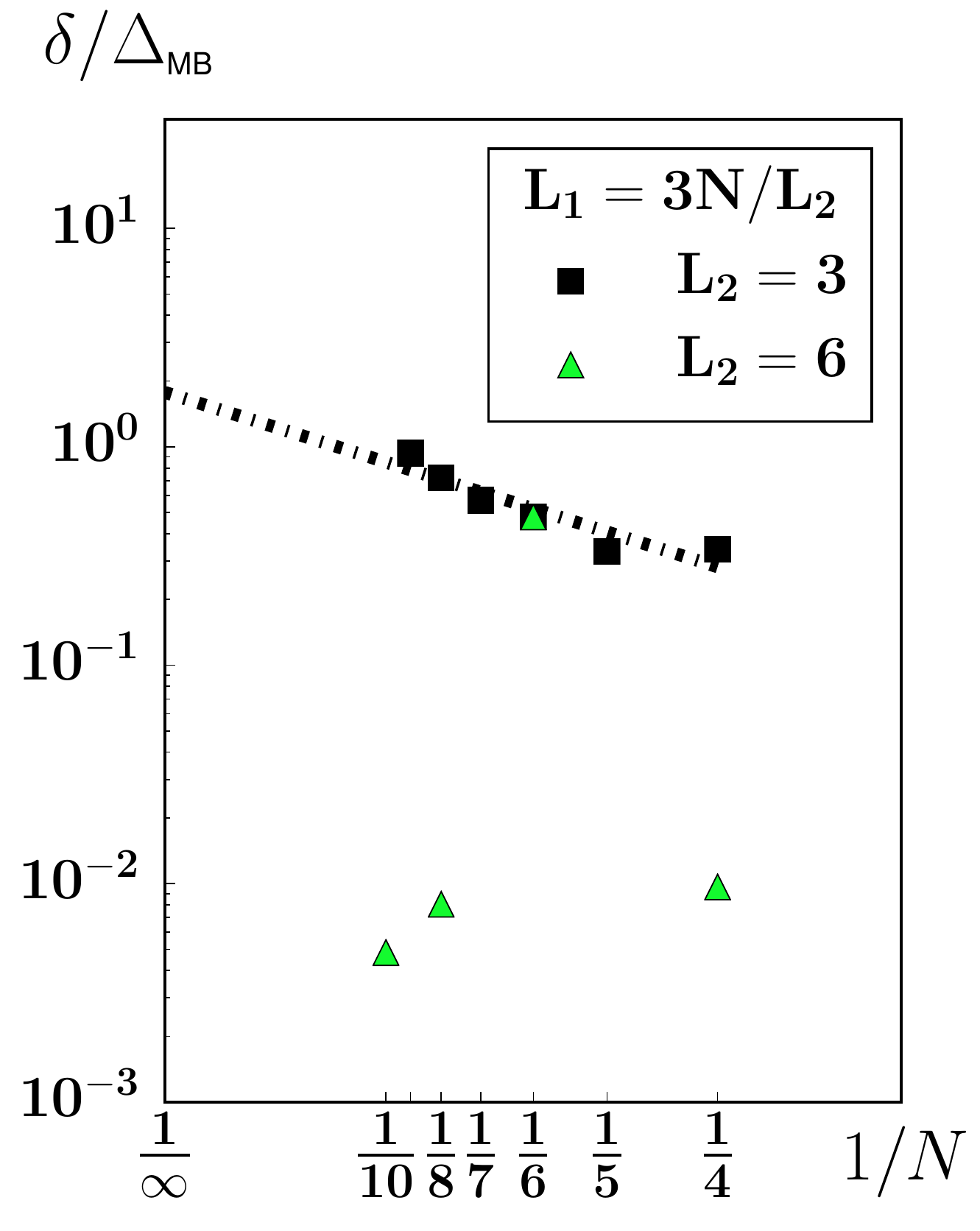}
\put(35,95){(b)}
\end{overpic}
\caption{
(color online). Thermodynamic extrapolation of the many-body energy gap $\deltamb$ (defined in the text) of $H_{\mbox{\tiny lb}}$ (\eq{eqInteractingHamiltonian}) at filling factor $\nu=1/3$ for different aspect ratios of the simulation grid and for $\tilde g/2 \pi = 1$. Dashed lines show linear extrapolations, where helpful:
(a) Many-body energy gap, $\deltamb$.
(b) Ratio of the energy spread, $\delta$ (defined in the text) to the energy gap $\deltamb$.}
\label{figGapScaling}
\end{minipage}
\end{figure}
}

\newcommand{\placeFigHalfFIllingGroundState}
{
\begin{figure}[t]
\centering
\includegraphics[width=\columnwidth]{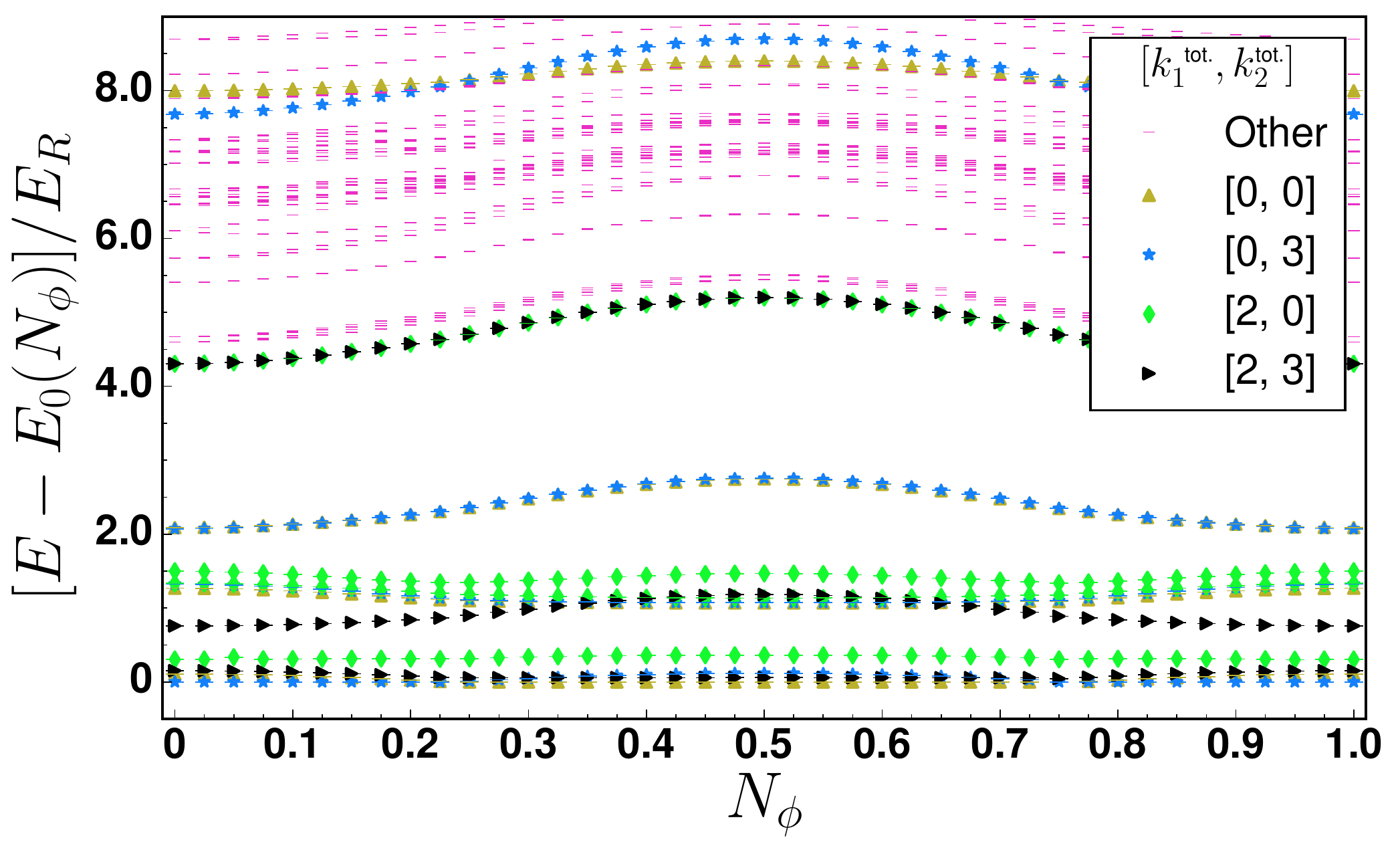}
\caption{
(color online).  Energy spectrum of $H_{\mbox{\tiny lb}}$ (\eq{eqInteractingHamiltonian}) for filling factor $\nu=1/2$ of the lowest OFL band ($N=12$ on a $4\times6$ grid) with $V_0/E_R=2.0$ and $\tilde g/2\pi = 1$, as a function the number, $N_{\phi}$, of flux quanta inserted along the $\boldsymbol \kappa'_2$ direction ($N_{\phi}=1$ corresponds to $\deltaky = \kappa'_2/L_2$). For clarity, for each $N_{\phi}$,  the minimum energy, $E_0 (N_{\phi})$, has been subtracted off. No simple spectral flow is observed. 
}
\label{figHalfFillingGroundState}
\end{figure}
}

\newcommand{\placeFigHalfFIllingCorrelations}
{
\begin{figure}[t]
\centering
\hspace{-2mm}
\begin{minipage}{\columnwidth}
\hspace{-2mm}
\begin{overpic}[width=0.495\columnwidth]{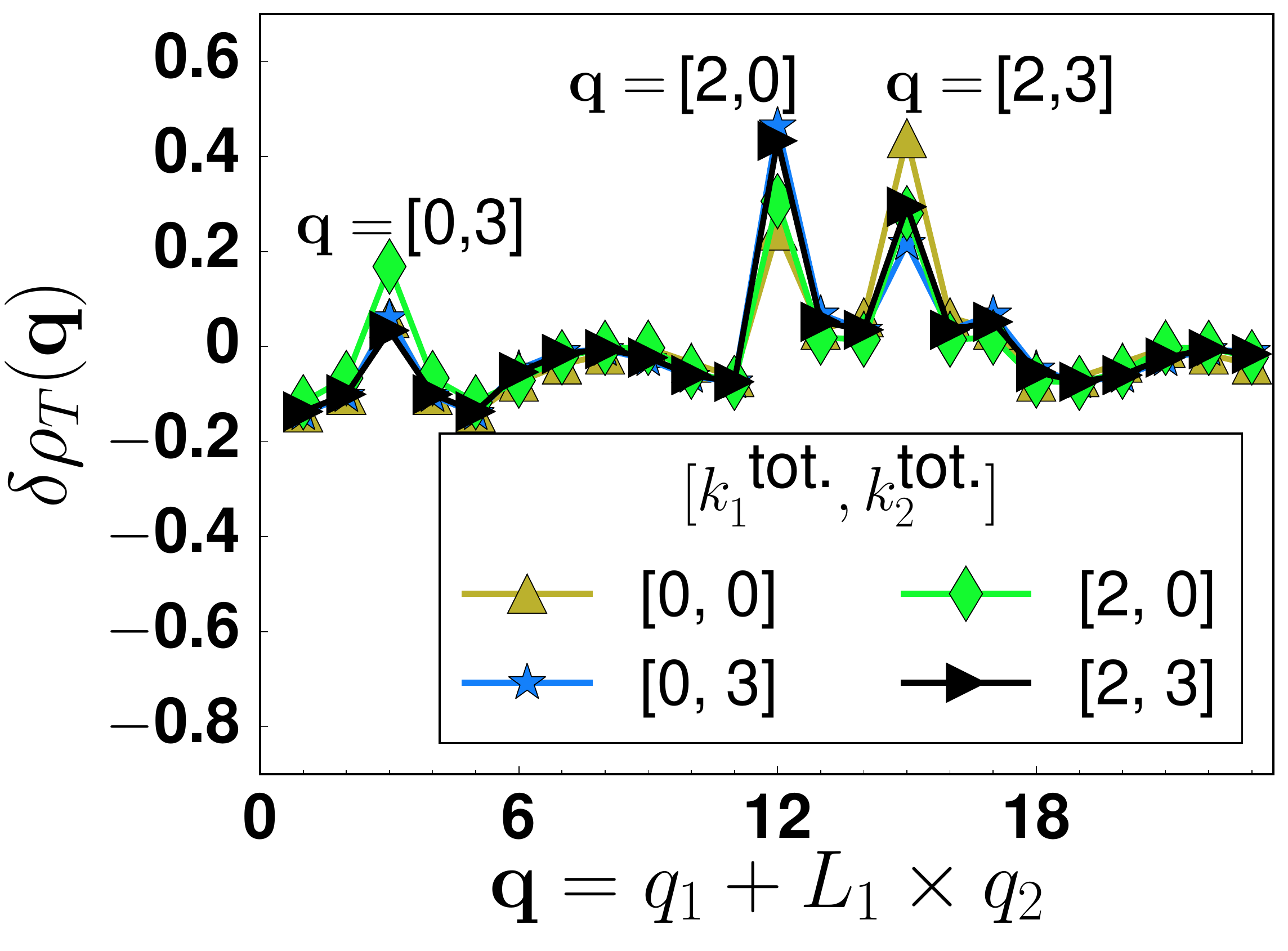}
\put(5,0){(a)}
\end{overpic}
\begin{overpic}[width=0.505\columnwidth]{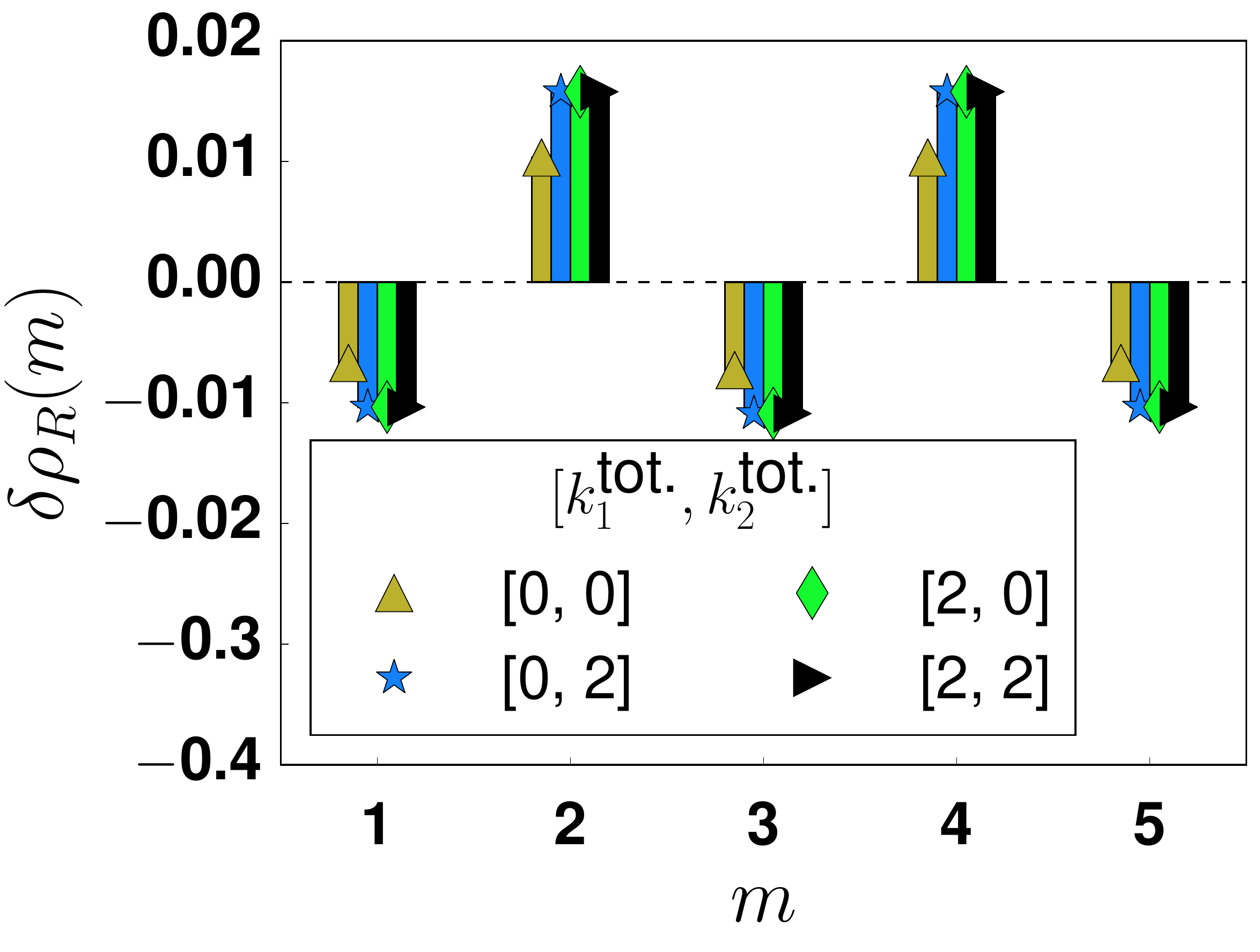}
\put(5,0){(b)}
\end{overpic}
\end{minipage}
\caption{
(color online).  Character of the low lying states of $H_{\mbox{\tiny lb}}$ (\eq{eqInteractingHamiltonian}) for filling factor $\nu=1/2$ of the lowest OFL band with $V_0/E_R=2.0$ and $\tilde g/2\pi = 1$:  (a) Representative fluctuations $\delta\rhoT ({\mathbf q}) $ in the static structure factor $\rhoT ({\mathbf q}) $ [defined in the text] for the lowest lying states ($N=12$ on a $4\times6$ grid). Peaks with $\left|\delta\rhoT ({\mathbf q}) \right| \sim 1 $ here indicate translational symmetry breaking; (b) Representative fluctuations $\delta \rhoR(m)$ in rotational correlation function $\rhoR (m)$ [defined in the text] for the lowest lying states ($N=8$ on a $4\times4$ grid). $\left|\delta\rhoR (m) \right| \sim 1/N $ here indicates 6-fold rotational invariance. 
}
\label{figHalfFillingCorrelations}
\end{figure}
}

\newcommand{\placeFigLaughlinQuasiholes}
{
\begin{figure}[t]
\includegraphics[width=\columnwidth]{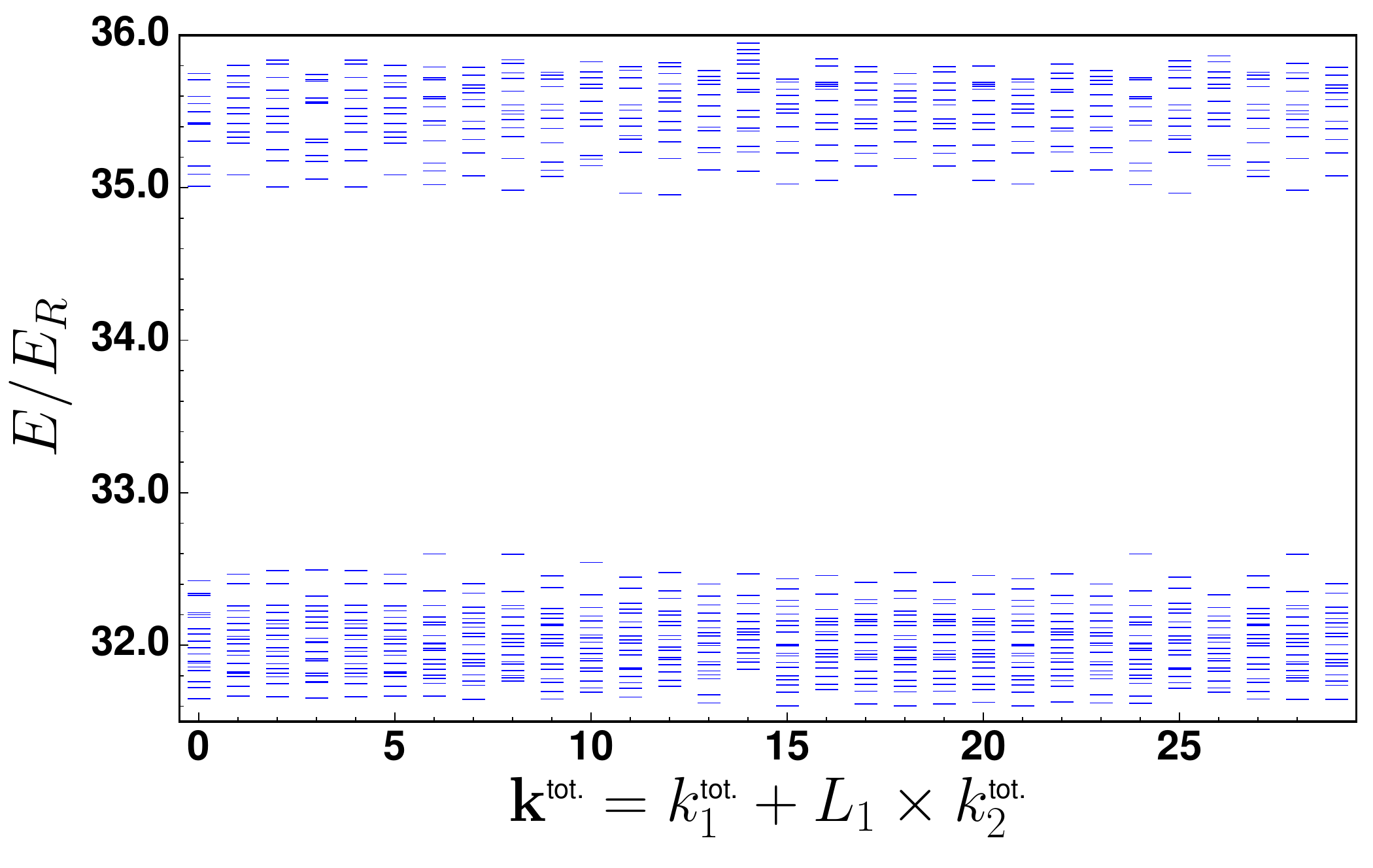}
\caption{
(color online). Energy spectrum of $H_{\mbox{\tiny lb}}$ (\eq{eqInteractingHamiltonian}) for $N=9$ on a $5\times6$ grid with $V_0/E_R=2.0$ and $\tilde g/2\pi = 1$ as a function of linear momentum sector $[k_1^{\mbox{ \tiny tot.}},k_2^{\mbox{\tiny tot.}}]$. The counting of quasi-hole states (energy levels below the gap) per sector precisely matches the analytic result given e.g. in \refr{Regnault2011}. 
}
\label{figQuasiHoleSpectrum}
\end{figure}
}

\begin{document}

%%%%%%  TITLE  %%%%%%%%%%%%%%%%%%%%%%%%%

\title{Ferromagnetic--nematic order and strongly correlated phases of fermions in optical flux lattices}

\author{Simon C. Davenport}
\author{Nigel R. Cooper}
\affiliation{T.C.M Group, Cavendish Laboratory, J.J. Thomson Avenue, Cambridge CB3 0HE, United Kingdom}

%% PACS

%% In 2015, APS will introduce a new classification scheme (taxonomy) for physics that will replace PACS.

%\pacs{
%67.85.Lm,   % Degenerate Fermi gases
%03.75.Ss,    % Degenerate Fermi gases
%03.65.Vf,      % Phases: geometric; dynamic or topological
%73.22.Gk     % Broken symmetry phases
%}

\date{\today}

%%%%%%  ABSTRACT  %%%%%%%%%%%%%%%%%%%%%%

\begin{abstract}

We study a model of a 2D ultracold atomic gas subject to an ``optical flux lattice'': a laser configuration where Raman-dressed atoms experience a strong artificial magnetic field. This leads to a bandstructure of narrow energy bands with non-zero Chern numbers. We consider the case of two-level (spin-$1/2$) fermionic atoms in this lattice, interacting via a repulsive $s$-wave contact interaction. Atoms restricted to the lowest band are described by an effective model of spinless fermions with interactions that couple states in a momentum-dependent manner across the Brillouin zone; a consequence of the Raman dressing of the two spin states. We present the results of  detailed exact diagonalization studies of the many-body states for a range of filling factors, $\nu$. First, we present evidence for the existence of a phase with coupled ferromagnetic--nematic ordering, which was previously suggested by a mean-field analysis. Second, we present evidence indicating the presence of a Laughlin-like fractional quantum Hall state occurring at filling factor $\nu = 1/3$. Finally, we observe a charge density wave state at $\nu=1/2$, which we are able to cleanly distinguish from the Laughlin-like state by its translational symmetry breaking and relatively small participation ratio. 

%% Subject Areas
\vspace{1em}
\noindent
Subject Areas: Condensed Matter Physics --- Strongly Correlated Materials; Atomic and Molecular Physics.

\end{abstract}

\maketitle

%%%%%%  INTRODUCTION   %%%%%%%%%%%%%%%%%

There is significant interest currently in experimental realizations of both topological phases of matter and itinerant ferromagnetism in ultracold atomic gases; \cite{Duine2005,Conduit2009,Conduit2010,Pilati2010,Chang2011} notwithstanding that, to date, topological phases have not yet been observed, \cite{Cooper2008,Bloch2008,Lin2009,Aidelsburger2011} while ferromagnetic behaviour has proved challenging to detect. \cite{Jo2009,Pekker2011,Sanner2012} In this paper we present evidence demonstrating the promise of ``optical flux lattices'' (OFLs) --- optical lattices that generate artificial magnetic fields via coherent Raman coupling of the internal states --- to advance both aims. \cite{Cooper2011,Cooper2011b} We do so by presenting the first results from exact diagonalization calculations of strongly interacting fermions in optical flux lattices. (Previous studies have considered bosonic atoms. \cite{cooperdalibard2013,Sterdyniak2015})

The key feature that leads to the usefulness of OFLs in stabilizing both ferromagnetic phases and topologically-ordered phases is that, throughout a large region of their parameter space, they exhibit narrow, almost flat bands with non-trivial band topology (non-zero Chern number), due to the presence of a net non-zero artificial magnetic field. \cite{Cooper2011}

From the perspective of ferromagnetic ordering, the band flattening leads to an enhanced density of states without exponentially localized Wannier orbitals, \cite{Thouless1984} allowing the Stoner instability to a ferromagnetic phase to occur at a value of interaction strength that is much reduced from the value required in free space. This has been demonstrated within a mean-field treatment of the interactions, \cite{Baur2012} which also shows that, owing to the Raman coupling of the spin states, the ordered phase has combined ferromagnetic--nematic order, with broken rotational symmetry of both the spin and orbital motion. \cite{Baur2012} Given that this phase appears in the strongly interacting regime, it is important to question the extent to which mean-field theory accurately describes the system. Our exact diagonalization studies provide clear evidence for the presence of ferromagnetism and the associated rotational symmetry breaking. Indeed, we find that the ferromagnetic--nematic phase appears at an even smaller interaction strength than in mean-field theory. The interaction strength is critical to practical implementations because it greatly influences the rate of three-body loss processes, which amongst other challenges, impede experimental progress. \cite{Pekker2011,Sanner2012} That the interaction strength is much reduced from its free-space value by the OFL can very significantly reduce these loss processes.

From the perspective of topological phases, the nearly flat and topological bands of the OFL make them ideal candidates in the search for topological fermionic phases analogous to the fractional quantum Hall (FQH) effect of electrons.  Here we shall provide compelling evidence demonstrating that indeed there is a strong FQH-like phase analogous to a Laughlin state at filling factor $\nu=1/3$ for interacting fermions in the lowest band.  There are a number of existing theoretical proposals to engineer fermionic FQH-like phases in band structures with non-trivial topology, for instance using a combination of precisely tuned nearest-neighbour and next-nearest-neighbour tunnelling and either on-site interactions,\cite{Tang2011,Neupert2011,Sheng2011,Regnault2011,Fei2011,Venderbos2012,Bergholtz2013} or dipolar interactions. \cite{Zeng2015} Such phases are often termed fractional Chern insulators (FCI). The presence of a Laughlin-like state in an OFL is intriguing in this context because the interactions have a rather different form from what has been considered previously. Although they arise from a 2-body $s$-wave contact interactions, once projected to the lowest band they acquire a non-local character with an unusual $\mathbf{k}$-dependent structure (a feature that originates from the optical dressing as we shall describe in more detail momentarily). Nevertheless, we find that the OFL's Laughlin-like state is surprisingly robust.

In addition to the aforementioned ferromagnetic--nematic phase and the FQH-like phase, we observe a highly incompressible state at $\nu=1/2$. Unlike at $\nu=1/3$, we show that this state breaks both translational and rotational symmetry, which allows us to clearly distinguish it as a charge density wave (CDW).

%%%%%%  BACKGROUND ON OFL MODEL  %%%%%%%

\subsection*{Model}

An optical lattice is formed by the interference pattern of linearly polarized in-plane laser beams with common wave vector, $\kappa$. We study a hexagonal lattice formed by three such beams at orientations differing by $2\pi/3$, which are described by wave vectors $\boldsymbol \kappa_1= -\kappa/2 (\sqrt3,1)$, $\boldsymbol \kappa_2= \kappa/2 (\sqrt3,-1)$ and $\boldsymbol \kappa_3=  \kappa (0,1)$. In order to engineer a spin-like degree of freedom one exploits the fact that atoms possess hyperfine atomic levels (which can be split by the Zeeman effect). For simplicity we focus on a 2-level system with angular momentum $F=1/2$. Practical implementations might have higher values of $F$, but the key features at $F=1/2$ should remain. \cite{Baur2012} The 2-level system is coherently Raman coupled, or dressed, by a fourth, circularly polarized light beam oriented perpendicular to the 2D plane of the lattice. This configuration leads to a 2-component Hamiltonian (in the notation of \refr{Baur2012})
\begin{equation}
\label{eqSingleParticleHamiltonian}
\hat{H}_0 = \frac{{\mathbf p^2}}{2M} \hat{\mathbb{1}} + V_{\mbox{\tiny sc}} ({\mathbf r}) \hat{\mathbb{1}} + \hat{{\boldsymbol \sigma}} \cdot {\mathbf B} ({\mathbf r}).
\end{equation}
The optical lattice is described by a scalar potential
\begin{equation}
\label{eqScalarPotential}
V_{\mbox{\tiny sc}} = V_0 \left[3\cos^2 (\theta) - 1\right] \sum_j \cos({\boldsymbol \kappa'_j \cdot \mathbf r}),
\end{equation}
and the Raman coupling is described by a scalar product of a vector of Pauli matrices $\hat{{\boldsymbol \sigma}}$ with a local vector field ${\mathbf B}$, whose components are
\begin{align}
& B_z = \sqrt{3} V_0 \sin^2 (\theta) {\sum_j \sin({\boldsymbol \kappa'_j \cdot \mathbf r})}, \\
& B_x + iB_y = \epsilon V_0 \cos(\theta) \sum_j \exp^{-i {\boldsymbol \kappa'_j \cdot \mathbf r}},
\end{align}
where $\boldsymbol \kappa'_1 = \boldsymbol \kappa_2 -\boldsymbol  \kappa_3$, $ \boldsymbol \kappa'_2= \boldsymbol \kappa_3 - \boldsymbol \kappa_1$ and $\boldsymbol \kappa'_3= \boldsymbol \kappa_1 - \boldsymbol \kappa_2$. \cite{Note1}
The tunable model parameters are the lattice depth, $V_0$, the linear polarization angle $\theta$ of the beams forming the optical lattice, and the amplitude ratio $\epsilon$ between the in-pane beams and the fourth, perpendicular beam. 
To reduce the number of free parameters, and for direct comparison with the existing mean-field theory results, we fix $\epsilon=0.4$ and $\theta=0.3$. \cite{Baur2012} We shall, however, consider the dependence of the properties on lattice depth $V_0$, as compared to the recoil energy, $E_R = \hbar^2 \kappa^2 / {2M}$.

The spectrum of \eq{eqSingleParticleHamiltonian} exhibits a set of energy bands with topological character.  In particular, for any non-zero $V_0/E_R$, the lowest energy band has unit Chern number.  It is analogous to the lowest Landau level in a system in which there is one flux quantum per real space unit cell. \cite{Cooper2011b} Thus, the ``quantum Hall'' filling factor, $\nu$ --- defined as the ratio of 2D particle density to magnetic flux density --- is equivalent to the band filling factor, i.e. the number of atoms per unit cell. 
The dispersion of this lowest energy band $\epsilon({\mathbf k})$ varies with $V_0/E_R$, having a minimum at around $V_0/E_R = 2. $\cite{Cooper2011b}

For our numerical studies, we construct the energy eigenstates of \eq{eqSingleParticleHamiltonian}  by proposing a set of Bloch wave functions on a $L_1 \times L_2$ $k$-space grid $\left\{{\mathbf K}\right\} = \boldsymbol \kappa'_1 l_1/L_1-\boldsymbol  \kappa'_2 l_2/L_2$ for $l_j =0,\ldots L_j-1$ with $j=1,2$, and a set of reciprocal lattice vectors $\left\{{\mathbf G}\right\}$:
\begin{equation}
\label{eqSingleParticleBlochStates}
\phi(\mathbf r)^{n}_{\mathbf k, \sigma} = \sum_{\mathbf g \in \left\{{\mathbf G }\right\} } A^{{n, \mathbf g}}_{{\mathbf k},\sigma} \exp^{i({\mathbf k+\mathbf g}) \cdot {\mathbf r}}, 
\end{equation}
where $n$ is a band index, $\sigma$ is a spin index and $ {\mathbf k} \in \left\{{\mathbf K}\right\}$.  
The Raman coupling induces a ${\mathbf k}$-dependent mixture of spin-up and spin-down character, $S_{\mathbf k} \in \left[ -1/2,1/2\right]$, in the single-particle states as a function of the amplitudes $A^{{n, \mathbf g}}_{{\mathbf k},\sigma} $. In the hexagonal Brillouin zone of the lowest band, single-particle states close to the $K$/$K'$ points obtain a largely spin-up/spin-down character. The interaction-driven ferromagnetic--nematic phase is, therefore, associated with a breaking of the original 6-fold symmetry of the Brillouin zone down to a 3-fold symmetry where states close to $K$/$K'$ are unevenly populated.

The simplest interactions that can be introduced into this model are $s$-wave contact interactions between spin-up and spin-down. We represent these by a two-dimensional delta-function interaction potential, of strength $\tilde g\hbar^2 /M$, setting the dimensionless interaction strength $\tilde g$. We consider repulsive interactions (positive $\tilde g$). In practical implementations, the magnitude and sign of these interactions can be tuned using a Feshbach resonance. \cite{Chin2010} The projection of that contact interaction to the lowest band produces an effective interaction between spinless fermions (again, owing to the Raman coupling). The effective Hamiltonian describing the lowest band is
\begin{equation}
\label{eqInteractingHamiltonian}
H_{\mbox{\tiny lb}} = \sum_{\mathbf k} {\boldsymbol \epsilon}_{\mathbf k} {c^\dagger}_{\mathbf k} c_{\mathbf k} + \frac{1}{2} \sum_{{\mathbf k}_1 {\mathbf k}_2 {\mathbf k}_3 {\mathbf k}_4} V_{{\mathbf k}_1 {\mathbf k}_2 {\mathbf k}_3 {\mathbf k}_4}  {c^\dagger}_{{\mathbf k}_1} {c^\dagger}_{{\mathbf k}_2} c_{{\mathbf k}_3} c_{{\mathbf k}_4},
\end{equation}
with
\begin{equation}
\label{eqVkkkk}
V_{{\mathbf k}_1 {\mathbf k}_2 {\mathbf k}_3 {\mathbf k}_4} =\frac{\tilde g\hbar^2} {M} \int d^2 {\mathbf r} \sum_{\sigma} \phi^{*}_{{\mathbf k}_1 \sigma} ({\mathbf r}) \phi^{*}_{{\mathbf k}_2 \bar\sigma} ({\mathbf r}) \phi_{{\mathbf k}_3 \bar\sigma} ({\mathbf r}) \phi_{{\mathbf k}_4\sigma} ({\mathbf r}),
\end{equation}
where $\bar{\sigma}$ denotes a spin flip, and we have suppressed the band index $n$ since we have projected to the lowest band.

We shall present detailed analyses of the ground states and low-energy spectra of $H_{\mbox{\tiny lb}}$ for a range of filling factors, lattice depths and interaction strengths. We study $N$ particles on the simulation grid $\left\{{\mathbf K}\right\}$. One is free to change the offset $\deltakx$ or $\deltaky$ between the grid $\left\{{\mathbf K}\right\}$ and the underlying Brillouin zone along directions $\boldsymbol \kappa'_1$ and $-\boldsymbol \kappa'_2$ respectively. (This freedom is equivalent to flux insertion in the two cycles of the torus formed by the periodic boundary conditions.) By varying these offsets we can, for example, check the extent to which our results are influenced by finite-size effects. 

%%%%%%  EVIDENCE FOR FERROMAGNETISM  %%%

\subsection*{Ferromagnetic--nematic order}

First, we shall identify the presence of a ferromagnetic--nematic phase in this model at low filling factors. That two forms of ordering appear in this phase is a consequence of the spin-orbit coupling inherent in the Raman coupling.

To quantify the appearance of ferromagnetic--nematic order, we measure the normalized density-density correlation function weighted by the spin character $S_{\mathbf k}$ at each point in $k$-space, given by 
\begin{equation}
\label{eqSusceptibility}
\rhoSpin = \frac{4}{N^2} \sum_{\mathbf k ,\mathbf k'} \left\langle {S_{\mathbf k'} c^\dagger}_{\mathbf k'} c_{\mathbf k'} S_{\mathbf k} {c^\dagger}_{\mathbf k} c_{\mathbf k} \right\rangle.
\end{equation}
For complete ferromagnetic order, with $S_{\mathbf k}  = + 1/2$ or $S_{\mathbf k}  =-1/2$, one would have $\sum_{\mathbf k} {S_{\mathbf k} c^\dagger}_{\mathbf k} c_{\mathbf k} = N/2$ and $\rho_s=1$. 
Note, however, that the spin $S_{\mathbf k}$ has some dependence on the lattice depth parameter, $V_0$, and $|S_{\mathbf k}|<1/2$.

In order to diagnose the presence of (or lack of) the associated nematic order, we study rotational symmetry breaking in $k$-space.
 We calculate a rotational density-density  function that measures commensurablility under changes in angular momentum  $m$:
\begin{equation}
\rhoR(m) = \frac{1}{6 N^2} \sum_{\mathbf k, \alpha} \exp^{i m \alpha} \left\langle {c^\dagger_{\hat R_{\alpha} \mathbf k} c_{\hat R_{\alpha} \mathbf k} c^\dagger_{\mathbf k} c_{\mathbf k} }\right\rangle \, 
\label{eqCorrelationFunctionsR},
\end{equation}
where $\hat R_{\alpha}$ are operations to rotate the k-states by angle $\alpha$. In order to elicit signatures indicating breaking of the original 6-fold symmetry we select $\alpha=n\pi/3$ and  $ m,n \in \{0$--$5\}$. $\rhoR (m)$ is only defined when the boundary conditions of the simulation grid themselves obey 6-fold rotational symmetry. The signature of $k$-space rotational symmetry breaking is that fluctuations 
\begin{equation}
\delta\rhoR (m) =  \frac{\rhoR (m) - \langle\rhoR (m) \rangle}{\langle\rhoR (m) \rangle}
\label{eqFluctuationsR}
\end{equation}
in $\rhoR (m)$ satisfy $\left| \delta\rhoR (m) \right| \sim 1$ for some $m$. Note that we choose a convention where the mean value $ \langle\rhoR (m) \rangle$ does not include $\rhoR (0)$ but does include $\rhoR (1)$ to $\rhoR (5)$. 

In \fig{figMagneticPhaseDiagram} we plot a map of the spin-weighted density-density function $\rhoSpin$ (\fig{figMagneticPhaseDiagram}a) and a map of the fluctuation in the rotational correlation function $\delta\rhoR (m=3)$  (\fig{figMagneticPhaseDiagram}b) for the ground state at filling factor $\nu=1/4$ for a range of interaction strength $\tilde g$ and lattice depth $V_0$. The plots are for particular values of the offsets $\deltakx$ and $\deltaky$ that allow $\left\{{\mathbf K}\right\}$ to have 6-fold rotational symmetry.  We have also checked that the values of $\rhoSpin$ are largely independent of the boundary conditions. 
Non-zero $\rhoSpin$ indicates a magnetic ground state and $\delta\rhoR (3) \sim 1$  indicates that the 6-fold rotational symmetry is broken to 3-fold.

From these results we conclude that evidence from these small systems supports the mean-field theory prediction of interaction-induced ferromagnetic--nematic order. Our results differ from mean-field theory in that they typically predict a partially polarized, rather than fully polarized ferromagnet. Also, the value of $\tilde g$ necessary to see some degree of ferromagnetism is significantly smaller than the mean-field theory prediction (see \refr{Baur2012}). Indeed, from \fig{figMagneticPhaseDiagram} it would appear that  relatively weak interactions and small $V_0$ are sufficient for ferromagnetism to occur. In particular, for non-zero $V_0/E_R$, the ferromagnetic--nematic phase appears in regimes of interaction $\tilde g$ that are as much as ten times smaller than the value $\tilde g = 2\pi$ of the Stoner instability for 2D itinerant fermions in free space. This  observation is very encouraging from an experimental perspective, as it indicates that ferromagnetic ordering can be found in regimes of interactions far from the Feshbach resonance where three-body losses are much suppressed.

% Put the magnetization map figure here
\placeFigMagnetisation

%%%%%%  STRONGLY CORRELATED PHASES

\subsection*{Strongly correlated phases}

Next we turn our attention to the identification of strongly correlated phases in the OFL. In \fig{figIncompressibility} we plot the inverse compressibility $\beta^{-1} (\nu)$ for the ground state of the lowest OFL band as a function of filling factor $\nu$ when the model is in its narrow band regime. This quantity is given by estimating the second derivative with respect to particle number $N$ of the ground state energy $E_{\mbox{\tiny GS}}^{N}$:
\begin{equation}
\label{eqIncompressibility}
\beta^{-1} (\nu=N/[L_1 L_2]) = \frac{1}{2 E_R} \left[ E_{\mbox{\tiny GS}}^{N+2} + E_{\mbox{\tiny GS}}^{N-2} - 2E_{\mbox{\tiny GS}}^{N} \right].
\end{equation}
The most prominent features in \fig{figIncompressibility} are peaks in the inverse compressibility at filling factors $\nu = 1/2$ and $\nu=1/3$ that occur only once interactions are introduced ($\tilde g>0$).

% Put the inverse compressibility figure here
\placeFigCompressibility

There are a number of possibilities that would be consistent with these inverse compressibility signatures: For instance a FQH-like state, a CDW, or a striped phase of the so-called ``thin-torus pattern'' (that is, stripes of a FQH-like ground state that would occur in the long, thin torus limit of the model). To assist in distinguishing between these different candidates we calculate the ``fractional participation ratio'',
\begin{equation}
\label{eqParticipationRatio}
P = \frac{\left(  \sum_{i=1}^{D} \left| C_i \right|^2 \right)^2}{D \sum_{i=1}^{D} \left| C_i \right|^4},
\end{equation}
where the sum is over $D$ possible Fock states $i= 1,\ldots, D$ with amplitudes $C_i$. \cite{Bell1970,Thouless1974} $P$ provides an estimate of the fraction of Fock states that significantly contribute to a many-body wave function. In \fig{figParticipationRatio} we show the fractional participation ratio for the ground state at filling factor $\nu=1/3$. For the set of model parameters used in this calculation, the OFL exhibits a narrow band when $V_0/E_R \approx 1.75$--$3.0$. We observe that $P$ is close to $0.33$ in the narrow band regime, for sufficiently strong interactions. This large value indicates that the ground state at $\nu=1/3$ is strongly correlated and cannot be well represented by a small number of Slater determinants. On the other hand, we find that the fractional participation ratio is much smaller for the ground states at $\nu=1/2$ and $\nu=1/4$, taking values on the order of $P=0.01$ (for a $4\times 6$ system with $N=12$) and $P=0.001$ (for a $6\times 6$ system with $N=9$), respectively. Therefore neither are strongly correlated. The result that  $\nu=1/4$ is not a strongly correlated phase is consistent with our conclusion (above) that the ferromagnetic--nematic ground state that we have identified here using exact methods is close to its mean-field description. 

% Put the participation ratio figure here
\placeFigParticipationRatio

%%%%%%  LAUGHLIN STATE	%%%%%%%%%%%%%%%%%
\subsubsection*{FQH-like state at $\nu=1/3$}

We shall now characterize the $\nu=1/3$ state in some more detail. First we shall investigate the effect of insertion of magnetic flux quanta $N_\phi$ through one cycle of the torus. The addition of a single flux quantum, $N_\phi=1$, is simulated by changing the boundary condition $\deltaky$ by precisely one lattice spacing, $\boldsymbol \kappa'_2/L_2$ (or similarly for the $\boldsymbol \kappa'_1$ direction). This procedure also shifts the spectrum with the same periodicity due to the underlying OFL band structure. In \fig{figLaughlinGroundState} we plot the energy spectrum of the $5\times6$ system as a function of $N_{\phi}$. Owing to the periodicity of the OFL in real space, the eigenstates of $H_{\mbox{\tiny lb}}$ can be subdivided into different linear momentum sectors $[\ktot_1,\ktot_2]$ along the two grid directions in $k$-space. We observe a 3-fold degenerate ground state manifold in the sectors expected for the Laughlin state. Once the underlying variation is removed---by subtracting off the mean energy of the ground state manifold $E_0(N_{\phi})$---we clearly observe the characteristic spectral flow pattern of the Laughlin state, whereby the ground states only return to their original ordering after the insertion of precisely 3 flux quanta (\fig{figLaughlinGroundState} inset).

In \fig{figGapScaling}a we plot the finite-size scaling of the many-body energy gap $\deltamb$ for the $\nu=1/3$ ground state. In \fig{figGapScaling}b we plot the finite-size scaling of the ratio of the many-body energy gap $\deltamb$ to the width $\delta$ of the ground state manifold. These values are estimated by the minimum possible $\deltamb$ and the maximum possible $\delta$ for a range of offsets $\deltakx$ and $\deltaky$. With the evidence available we conclude that the energy gap remains non-zero in the thermodynamic limit. For the $L_y=6$ data set we observe that $\deltamb /\delta$ becomes negligible for larger $N$. In the ``thin torus'' case ($L_y=3$), however, we observe that the ratio $\deltamb /\delta$ tends to a non-negligible value. This behaviour is similar to that reported for FCI models. \cite{Regnault2011}

The spectrum of quasi-hole states can be obtained by removing particles from the $1/3$ state (for instance, $N=9$ in a $5\times6$ system, see \fig{figQuasiHoleSpectrum}) and we find that the counting of quasi-hole states per linear momentum sector precisely matches the analytic formula given in \refr{Regnault2011}. 

% Put the Laughlin ground state figure here
\placeFigLaughlinGroundState

% Put the gap extrapolation figure here
\placeFigGapExtrapolation

% Put the Laughlin quasihole spectrum figure here
\placeFigLaughlinQuasiholes

% Put the Laughlin ground state correlations figure here
\placeFigLaughlinCorrelations

To  rule out the possibility that the $\nu=1/3$ state could be a CDW or striped phase, we look for the presence or absence of translational symmetry breaking
by computing the static structure factor
\begin{equation}
\rhoT(\mathbf q) = \frac{1}{N} \sum_{\mathbf k ,\mathbf k'} \left\langle {c^\dagger}_{\mathbf k'-\mathbf q} c_{\mathbf k'} {c^\dagger}_{\mathbf k+\mathbf q} c_{\mathbf k} \right\rangle, 
\label{eqCorrelationFunctionsT}
\end{equation}
where ${\mathbf q} \in \left\{{\mathbf K}\right\}$. As with the rotational correlation function, we can identify translational symmetry breaking by analysing the order of magnitude of fluctuations from the mean value 
\begin{equation}
\delta\rhoT ({\mathbf q}) =  \frac{\rhoT ({\mathbf q}) - \langle\rhoT ({\mathbf q}) \rangle}{\langle\rhoT ({\mathbf q}) \rangle},
\label{eqFluctuationsT}
\end{equation}
where again we choose the convention to disregard $\rhoT(0)$ from the calculation of  the mean. CDW or striped states break translational symmetry and will satisfy $\left| \delta\rhoT ({\mathbf q}) \right| \sim 1$ for some ${\mathbf q}$, but FQH-like states do not. FQH-like states should also be rotationally invariant, when possible on a lattice, owing to their uniform density. 

In \fig{figLaughlinCorrelations} we plot the fluctuations in the static structure factor ($\delta\rhoT$) and also the rotational correlation function ($\delta\rhoR)$ for the ground state at $\nu=1/3$. We clearly observe that it breaks neither translational nor (6-fold) rotational symmetry in $k$-space. The structure of the fluctuations is associated with the properties of the quantum Hall liquid (see e.g. \refr{Girvin1986,Read2011} for a discussion of the sub-structure of the static structure factor for a FQH state). Our study is restricted to very small systems and, as such, we cannot resolve that sub-structure in detail. 

% Put the nu=1/2 spectrum figure here
\placeFigHalfFIllingGroundState

%%%%%%  EVIDENCE FOR CHARGE DENSITY WAVE  AT HALF FILLING
\subsubsection*{CDW state at $\nu=1/2$}

% Put the nu=1/2 correlations figure here
\placeFigHalfFIllingCorrelations

Finally, we focus on the incompressible phase that appears at filling factor $\nu=1/2$. In \fig{figHalfFillingGroundState} we plot the energy spectrum of the $4\times6$ system as a function of the  number of flux, $N_\phi$ inserted through the $\boldsymbol \kappa'_2$ cycle of the torus. No simple spectral flow is observed and instead we see dispersion and mixing of a large number of low-lying energy levels. In \fig{figHalfFillingCorrelations}a we plot representative fluctuations in the static structure factor ($\delta\rhoT$) associated with these low lying states. Clear peaks are seen in $\delta\rhoT$ corresponding to momentum sectors satisfying $\mbox{mod}(q_j,L_j/2)=0$ for $j=1,2$. This character is also observed in the ground state and low lying states of the $\nu=1/2$ state in $4\times4$, $4\times8$ and $5\times6$ systems. It indicates the formation of a CDW ground state with a 
 unit cell that is doubled in size along both axes.  In \fig{figHalfFillingCorrelations}b we plot representative fluctuations in the rotational correlation function ($\delta\rhoR$) associated with these low lying states. We observe that they retain the 6-fold rotational symmetry in $k$-space. Our conclusion here is consistent with the above indication from the fractional participation ratio that $\nu=1/2$  is not strongly correlated.

%%%%%%  CONCLUSION  %%%%%%%%%%%%%%%%%%%%

\subsection*{Conclusion}

We have presented a significant body of evidence that  together supports the conclusions that interacting fermions in an OFL can exhibit: 1). A (partially polarized) ferromagnetic--nematic phase whose onset occurs at a much reduced interaction strength compared to the free space Stoner criterion; 2). A Laughlin-like state at $1/3$ filling; 3). A CDW state at $\nu=1/2$ in the narrow band regime.
We have employed a number of simple correlation functions with which to measure the symmetries of the model, and in particular to identify when these symmetries are broken, and how strongly. These correlation functions allow us to distinguish a very strong FQH-like state from CDW and striped states. Further we have demonstrated that the many-body participation ratio measure can also be used to support such distinctions. The specific evidence we have gathered is for a particular set of OFL parameters (fixed $\epsilon$ and $\theta$), however there is nothing special about our choice of these parameters and the narrow band regime pervades a large region of the OFL parameter space. Our results demonstrate that an OFL would be a fruitful approach with which to enhance ongoing efforts towards observing both ferromagnetic behaviour and more exotic phases of matter in cold atomic gases.

%%%%%%  ACKNOWLEDGEMENTS  %%%%%%%%%%%%%%

\vspace{0.5em}

\noindent \textbf{Acknowledgements:} We thank S. K. Baur and G. M\"{o}ller for helpful discussions. This research was supported by EPSRC grant EP/J017639/1. Statement of compliance with EPSRC policy framework on research data: All data accompanying this publication are directly available within the publication. We acknowledge use of the Darwin Supercomputer of the University of Cambridge High Performance Computing Service (http://www.hpc.cam.ac.uk/).

\newpage

%%%%%%  BIBLIOGRAPHY  %%%%%%%%%%%%%%%%%%

\bibliographystyle{prsty}
%\bibliography{myrefs}

\begin{thebibliography}{10}

\bibitem{Duine2005}
R.~A. Duine and A.~H. MacDonald, Phys. Rev. Lett. {\bf 95},  230403  (2005).

\bibitem{Conduit2009}
G.~J. Conduit, A.~G. Green, and B.~D. Simons, Phys. Rev. Lett. {\bf 103},
  207201  (2009).

\bibitem{Conduit2010}
G.~J. Conduit, Phys. Rev. A {\bf 82},  043604  (2010).

\bibitem{Pilati2010}
S. Pilati, G. Bertaina, S. Giorgini, and M. Troyer, Phys. Rev. Lett. {\bf 105},
   030405  (2010).

\bibitem{Chang2011}
S.-Y. Chang, M. Randeria, and N. Trivedi, Proc. Natl. Acad. Sci. {\bf 108},  51
   (2011).

\bibitem{Cooper2008}
N.~R. Cooper, Adv. Phys. {\bf 57},  539  (2008).

\bibitem{Bloch2008}
I. Bloch, J. Dalibard, and W. Zwerger, Rev. Mod. Phys. {\bf 80},  885  (2008).

\bibitem{Lin2009}
Y.-J. Lin, R.~L. Compton, K. Jiménez-García, J.~V. Porto, and I.~B. Spielman,
  Nature {\bf 462},  628  (2009).

\bibitem{Aidelsburger2011}
M. Aidelsburger, M. Atala, S. Nascimb\`ene, S. Trotzky, Y.-A. Chen, and I.
  Bloch, Phys. Rev. Lett. {\bf 107},  255301  (2011).

\bibitem{Jo2009}
G.-B. Jo {\it et~al.}, Science {\bf 325},  1521  (2009).

\bibitem{Pekker2011}
D. Pekker {\it et~al.}, Phys. Rev. Lett. {\bf 106},  050402  (2011).

\bibitem{Sanner2012}
C. Sanner, E.~J. Su, W. Huang, A. Keshet, J. Gillen, and W. Ketterle, Phys.
  Rev. Lett. {\bf 108},  240404  (2012).

\bibitem{Cooper2011}
N.~R. Cooper, Phys. Rev. Lett. {\bf 106},  175301  (2011).

\bibitem{Cooper2011b}
N.~R. Cooper and J. Dalibard, Europhys. Lett. {\bf 95},  66004  (2011).

\bibitem{cooperdalibard2013}
N.~R. Cooper and J. Dalibard, Phys. Rev. Lett. {\bf 110},  185301  (2013).

\bibitem{Sterdyniak2015}
A. Sterdyniak, B.~A. Bernevig, N.~R. Cooper, and N. Regnault, Phys. Rev. B {\bf
  91},  035115  (2015).

\bibitem{Thouless1984}
D.~J. Thouless, Journal of Physics C: Solid State Physics {\bf 17},  L325
  (1984).

\bibitem{Baur2012}
S.~K. Baur and N.~R. Cooper, Phys. Rev. Lett. {\bf 109},  265301  (2012).

\bibitem{Tang2011}
E. Tang, J.-W. Mei, and X.-G. Wen, Phys. Rev. Lett. {\bf 106},  236802  (2011).

\bibitem{Neupert2011}
T. Neupert, L. Santos, C. Chamon, and C. Mudry, Phys. Rev. Lett. {\bf 106},
  236804  (2011).

\bibitem{Sheng2011}
D. Sheng, Z.-C. Gu, K. Sun, and L. Sheng, Nat. Commun. {\bf 2},  389  (2011).

\bibitem{Regnault2011}
N. Regnault and B.~A. Bernevig, Phys. Rev. X {\bf 1},  021014  (2011).

\bibitem{Fei2011}
Y.-F. Wang, Z.-C. Gu, C.-D. Gong, and D.~N. Sheng, Phys. Rev. Lett. {\bf 107},
  146803  (2011).

\bibitem{Venderbos2012}
J.~W.~F. Venderbos, S. Kourtis, J. van~den Brink, and M. Daghofer, Phys. Rev.
  Lett. {\bf 108},  126405  (2012).

\bibitem{Bergholtz2013}
E.~J. Bergholtz and Z. Liu, Int. J. Mod. Phys. B {\bf 27},  1330017  (2013).

\bibitem{Zeng2015}
T.-S. Zeng and L. Yin, Phys. Rev. B {\bf 91},  075102  (2015).

\bibitem{Note1}
Note that, for convenience of notation here, the definitions of $
  \boldsymbol \kappa '_1$ and $ \boldsymbol \kappa '_3$ have been
  swapped compared to Ref.~\onlinecite{Baur2012}.

\bibitem{Chin2010}
C. Chin, R. Grimm, P. Julienne, and E. Tiesinga, Rev. Mod. Phys. {\bf 82},
  1225  (2010).

\bibitem{Bell1970}
R.~J. Bell and P. Dean, Discuss. Faraday Soc. {\bf 50},  55  (1970).

\bibitem{Thouless1974}
D.~J. Thouless, Phys. Rep. {\bf 13},  93   (1974).

\bibitem{Girvin1986}
S.~M. Girvin, A.~H. MacDonald, and P.~M. Platzman, Phys. Rev. B {\bf 33},  2481
   (1986).

\bibitem{Read2011}
N. Read and E.~H. Rezayi, Phys. Rev. B {\bf 84},  085316  (2011).

\end{thebibliography}

\end{document}